\documentclass[%
aip,jcp,amsmath,amssymb,twocolumn,reprint,superscriptaddress
]{revtex4-2}

\usepackage[utf8]{inputenc}
\usepackage[T1]{fontenc}
\usepackage{mathptmx}
\usepackage{etoolbox}

\usepackage{graphicx}
\usepackage{dcolumn}
\usepackage{soul}
\usepackage[dvipsnames]{xcolor}
\usepackage{mathptmx}
\usepackage{amssymb}
\usepackage{array}
\usepackage{amsmath}
\usepackage{bm}
\usepackage{setspace}
\usepackage{hyperref}

\usepackage[normalem]{ulem}
\definecolor{darkgreen}{rgb}{0.0, 0.8, 0.3}

\begin{document}

\title{Inferring interactions between active particles using harmonic traps}

\author{A. Compagnie}
\affiliation{Institut für Theoretische Physik II: Weiche Materie, Heinrich-Heine-Universität Düsseldorf, Düsseldorf, Germany}
\author{J. Mecke}
\affiliation{Institut für Theoretische Physik II: Weiche Materie, Heinrich-Heine-Universität Düsseldorf, Düsseldorf, Germany}
\author{I. Buttinoni}
\email{Ivo.Buttinoni@hhu.de}
\affiliation{Institut für 
Experimentelle Kolloidphysik, Heinrich-Heine-Universität Düsseldorf, Düsseldorf, Germany}
\author{H. Löwen}
\email{hlowen@hhu.de}
\affiliation{Institut für Theoretische Physik II: Weiche Materie, Heinrich-Heine-Universität Düsseldorf, Düsseldorf, Germany}


\begin{abstract}

Quantifying the interactions between active Janus particles remains a challenge that needs to be addressed to describe their collective behaviour. Inspired by a photonic force microscopy setup where two active Janus particles are confined in separate optical tweezers, we derive the steady-state probability distribution of each particle bearing signatures of the pair phoretic interaction, which can be generalised to a broader range of interactions. This approach allows us to infer interaction forces and torques from positional configurations without any knowledge of the particle orientations. Unraveling the distance-dependent interaction between active particles is essential to harnessing them as building blocks for advanced materials with dynamic properties.

\end{abstract}
\maketitle


\section{Introduction}

Activity, the ability to consume energy to cause directed motion~\cite{elgeti_physics_2015,bechinger_active_2016,marchetti_hydrodynamics_2013}, is observed at micro- and macro-scales, in both biological~\cite{dombrowski_self-concentration_2004,sokolov_concentration_2007,cavagna_bird_2014,ward_quorum_2008} and synthetic~\cite{dauchot_dynamics_2019,novkoski_graspion_2026,scholz_inertial_2018} systems. Depending on the mode of activity and interactions between the active agents, this can lead to many different collective behaviours, from motility-induced phase separation~\cite{cates_motility-induced_2015,bialke_microscopic_2013,buttinoni_dynamical_2013} to self-assembly, flocking, swarming, or clustering just to name a few~\cite{bar_self-propelled_2020,kaiser_active_2015,yan_reconfiguring_2016,liao_emergent_2021,hennes_self-induced_2014,dinelli_non-reciprocity_2023,knezevic_collective_2022,yu_pressure_2025,forgacs_transient_2023,adorjani_motility-induced_2024}.

Janus particles are a paradigmatic example of synthetic active agents at the microscale \cite{golestanian_propulsion_2005,howse_self-motile_2007,jiang_active_2010,gangwal_induced-charge_2008,buttinoni_active_2012}. Due to different properties of the two hemispheres, they are able to generate phoretic chemical or thermal gradients leading to local slip flows and consequent force-free self-propulsion~\cite{anderson_colloid_1989,tatulea-codrean_artificial_2018}. The resulting single-particle dynamics is then well described by the active Brownian particle (ABP) model where the colloid moves at constant speed\textemdash resulting from the phoretic surface slip\textemdash through the solvent while its orientation diffuses~\cite{howse_self-motile_2007}. %
However, the ABP description becomes inappropriate when interactions among several self-phoretic Janus colloids are important. The slip velocities entail contributions not only from the \emph{own} phoretic field, but also from the one of their neighbours. This effect, known as cross-phoresis~\cite{pohl_dynamic_2014,pohl_self-phoretic_2015,zottl_modeling_2023,illien_fuelled_2017,liebchen_phoretic_2017,scagliarini_unravelling_2020,nasouri_exact_2020}, can trigger variations in the magnitude and direction of the active velocity and introduce torques that reorient particles. Moreover, distortions of the surrounding solvent caused by a Janus colloid implies hydrodynamic interactions on its neighbours~\cite{ishikawa_hydrodynamic_2006,yoshinaga_hydrodynamic_2017,campbell_experimental_2019,mousavi_clustering_2019,zottl_hydrodynamics_2014}.
Understanding the influence and importance of all these interactions is challenging~\cite{liebchen_which_2019,liebchen_interactions_2022}. Analytical descriptions of these systems have been presented to predict the dynamics of the particles~\cite{sharifi-mood_pair_2016,kanso_phoretic_2019,nasouri_exact_2020}, and different methods have been designed to try to understand and quantify the effective interactions from observing the dynamics of such particles, compared to simulations~\cite{hem_learning_2025,bera_learning_2026} or experimentally~\cite{sharan_pair_2023,singh_pair_2024}.

\begin{figure*}[th!]
    \includegraphics[width=\linewidth]{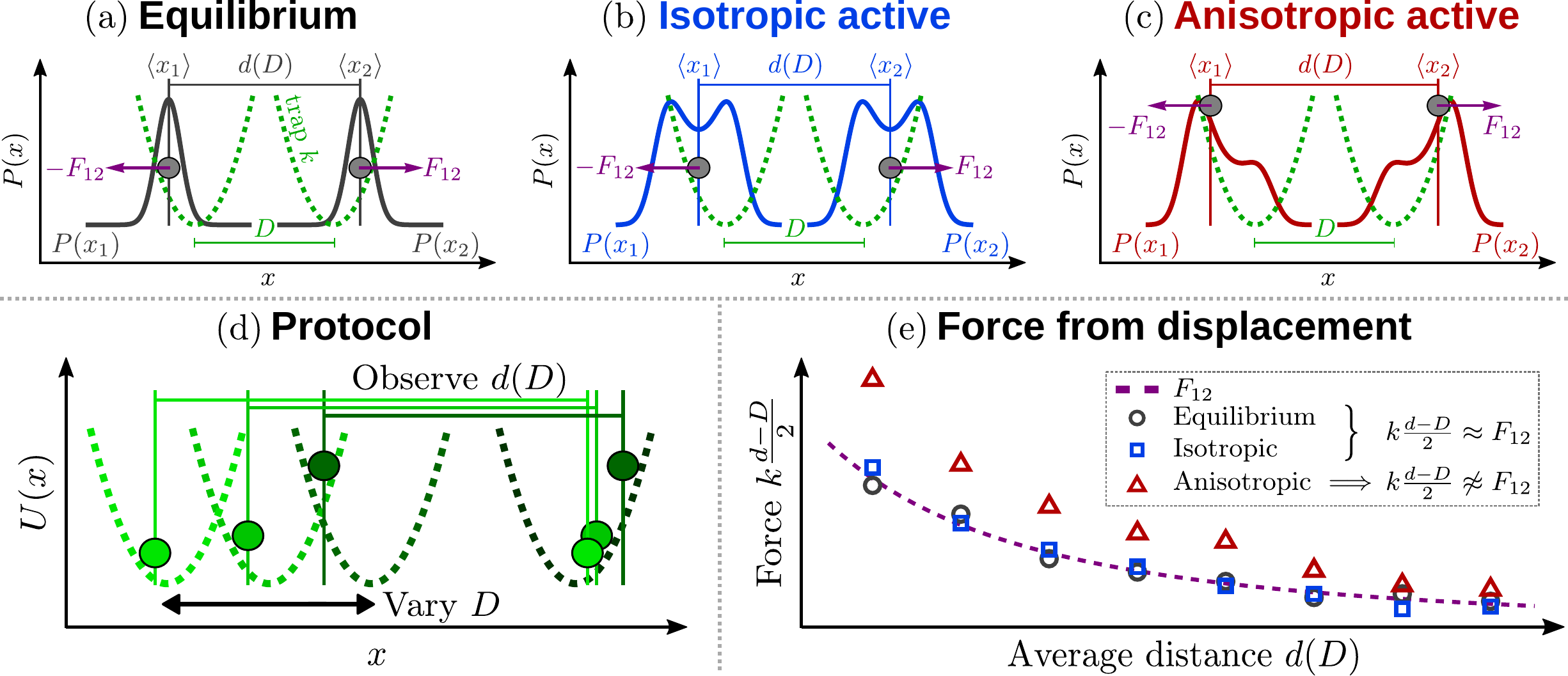}
    \caption{Schematic explanation of the effects of interactions to two particles in two different harmonic potential setup separated by a distance $D$. Representation of the probability distribution $P(x_i)$ of both particles interacting through a force $F_{12}(x)$, with $d=\langle x_2 \rangle - \langle x_1 \rangle$, (a) for passive particles in equilibrium, (b) for active Brownian particles, and (c) for active Brownian particles that also interact anisotropically. (d) The general protocol to understand how $F_{12}$ scales in space is to vary the distance $D$ between the traps and observe how the distance $d$ between the average positions responds, due to its direct link to the interactions. (e) For passive and isotropic active particles, the displacement of the average positions is directly proportional to the force $F_{12}$, contrarily to active particles with anisotropic interactions.}
    \label{fig:1}
\end{figure*}

Harmonic potentials constitute a reliable theoretical and experimental framework to measure pair interactions and model complex environments. Experimentally, such potentials can be realised by means of acoustic~\cite{takatori_acoustic_2016} or optical~\cite{ashkin_observation_1986,pesce_optical_2020} tweezers, which can be used to trap and move colloids. %
By studying the response of a trapped colloid due to the interaction with other colloids, one can infer the inter-particle interactions~\cite{meiners_direct_1999,buttinoni_mechanical_2021}. %
If two colloids are trapped in two optical traps of stiffness $k$ and relative distance $D$, the reciprocal pair-interaction force $F_{12}$ leads to a displacement of the colloids' average position (see Fig.~\ref{fig:1}a). To leading order, the probability distributions will be merely displaced and $F_{12}$ can be inferred from the displaced equilibrium distance between the colloids $d = |\langle x_1 \rangle - \langle x_2 \rangle|$ as $F_{12}(d)\approx k(d-D)/2$. 
However, extending similar setups to colloids under nonequilibrium conditions remains an open problem since equilibrium results cannot be employed to leapfrog the microscopic details. %
Only in the simplest case, of reciprocal body forces that neither depend on the colloids' distance nor their orientation, the bimodal marginal probability distribution of trapped ABPs~\cite{buttinoni_active_2022} is merely displaced (see Fig.~\ref{fig:1}b) such that we can apply $F_{12}(d)\approx k(d-D)/2$. %
However, under realistic conditions, distance dependent interactions break the isotropy of the ABP probability distribution. Moreover, (anti-)aligning torques, which are of paramount importance for phoretically interacting Janus colloids~\cite{pohl_dynamic_2014,pohl_self-phoretic_2015,liebchen_phoretic_2017,scagliarini_unravelling_2020}, break the isotropy of the colloidal angular degrees of freedom. Consequently, the active colloids' polarisation is non-zero: the orientational and translational degrees of freedom couple, and the positional probability distribution bears a signature of the underlying orientational anisotropy (see Fig.~\ref{fig:1}c) and $F_{12}(d)\not\approx k(d-D)/2$. %
Performing the same type of measurement for varying distances $D$ thus allows to measure the pair-interactions $F_{12}(r)$ (see Fig.~\ref{fig:1}d and e) only in equilibrium and nonequilibrium special cases. %

Despite the growing relevance of understanding the collective behaviour of interacting, aligning, or ``cooperating'' active colloidal systems, it remains a challenge to identify individual contributions in the pair interactions such as hydrodynamics, phoresis, or even van der Waals and double-layer forces, and sometimes unambiguous conclusions about the present interactions might even be impossible. %
In order to gain insight into the pair-interaction forces and torques between active Janus colloids, it is important to track the colloidal positions and orientations. However, unequal optical properties of the two hemispheres unavoidably trigger undesired side effects such as optical torques emanating from a refractive-index mismatch of the two materials~\cite{moyses_trochoidal_2016,aubret_metamachines_2021}. In order to isolate the phoretic contribution from the optical torques, the two hemispheres must be index-matched~\cite{buttinoni_active_2022}, which in turn prevents a straightforward tracking of the instantaneous orientation. 
Overcoming these challenges is important because the understanding of how the collective phenomena rely on the underlying interactions and the protocol of energy injection is not only of profound theoretical interest, but is also essential for the design of materials with tailored functionalities~\cite{xi_emergent_2024,ben_zion_morphological_2023,li_bioinspired_2023,zhang_nanomotor-based_2019,vyskocil_cancer_2020,urso_smart_2023}.

In this work, we derive the asymmetric signatures of the positional probability distribution stemming from the orientation-translation coupling of two interacting active Janus colloids held in two different optical traps. Although broadly applicable, our approach focuses specifically on cross-phoretic pair interactions leading to an additional propulsion and an orienting torque inducing a bias in the orientation that impacts the particle's position. %
We analytically express the probability distribution within the far-field approximation and quantify how the interactions alter steady-state properties. %
Our insights allow to infer the strength of the pair interaction forces and torques from the positional degrees of freedom alone. %
We thus extend the equilibrium force-displacement protocol $F_{12}(d)\approx k(d-D)/2$ to more general nonequilibrium conditions without the necessity of tracking the Janus colloids' orientational dynamics. %
Upon varying the distance between the traps, we can infer the distance dependence of the cross-phoretic interaction forces and torques. %
Our results are tested against Brownian dynamics simulations demonstrating excellent agreement. %
Our framework can be generalised to any setup of interacting active colloids with orientation-translation coupling such as (anti-)aligning active colloids and thus advances the fundamental understanding of potentially interactions between active colloids.

\section{Model} \label{sec:model}

\begin{figure}[t]
    \centering
    \includegraphics[width=\linewidth]{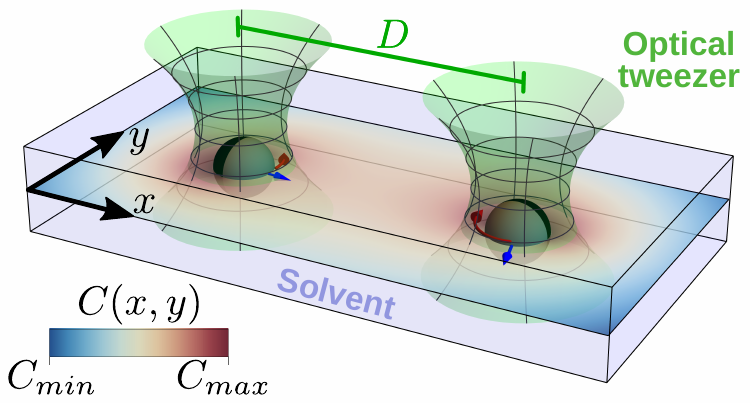}
    \caption{Schematic representation of the complete setup of the model, consisting of two Janus particles each trapped in different optical tweezers, moving in a 2D plane. The particles are affected by the phoretic field $C(x,y)$ through a force due to active self-phoresis and cross-phoresis (blue arrows), and through a torque due to cross-phoresis (red arrows). The forces are not necessarily aligned with the Janus particle orientation because of the cross-phoretic effects.}
    \label{fig:2}
\end{figure}

We model a system with two Janus particles trapped in two different optical tweezers, as represented in Fig.~\ref{fig:2}. We consider the motion of the particles in a 2D plane, where the particles are described by their position $\mathbf{r}_i$ and orientation using the angle $\theta_i$ or the unit vector $\hat{\mathbf{n}}_i = (\cos\theta_i,\sin\theta_i)^T$. 
Two optical tweezers, in which the two particles are trapped, are placed along the $x$ axis and separated by a distance $D$, as illustrated in green in Fig.~\ref{fig:2}. They are both modelled by harmonic potentials with a trap stiffness $k$~\cite{buttinoni_active_2022}. 

We first focus here on systems where the interaction that dominates comes from a 3D phoretic field generated by the Janus particles. We can consider the simplest form of a Janus particle, \emph{i.e.}, a perfect sphere with two halves of different physical or chemical properties, where one of them continuously acts as a source or a sink for a generic phoretic field $C(\mathbf{r})$, shown in the horizontal plane of Fig.~\ref{fig:2}. This can be the case, \emph{e.g.}, if one hemisphere catalyzes a chemical reaction and alters the chemical concentration field (diffusiophoresis) or generates heat and induces a temperature gradient (thermophoresis). %
The gradient of the joint phoretic field $C(\mathbf{r})$ around the Janus particles will induce a surface flow of the solvent, such that momentum conservation leads to force- and torque-free translation and rotation of the particles. %
This self-phoretic effect can be modelled within the ABP model, that is we assume a constant self-propulsion velocity $v_0$ and a diffusing orientation $\hat{\mathbf{n}}_i$ as a result from the co-moving stationary phoretic field around each colloid. %

In order to model cross-phoretic effects, \emph{i.e.}, propulsion of colloid $i$ due to influences on the phoretic field of colloid $j$, we assume the distance between the particles to be significantly greater than their size, such that we reach the far-field limit. In a surrounding 3D medium the phoretic field around each colloid can be approximately described as a radially symmetric point source or sink monopole in leading order for the field $C_j(\mathbf{r}_i) \approx c |\mathbf{r}_j-\mathbf{r}_i|^{-1}$. The constant $c$ depends on the diffusivity of the phoretic field and the strength of the source or sink. %
Within the far-field approximation, the phoretic thrust and torque are then obtained as $\mathbf{F}_i/\gamma = -\mu_t \mathbf{\nabla}C_j(|\mathbf{r}_i-\mathbf{r}_j|)$ and $\mathbf{T}_i/\zeta = -\mu_r \hat{\mathbf{n}}_i \times \mathbf{\nabla}C_j(|\mathbf{r}_i-\mathbf{r}_j|)/\sigma$, where $\gamma$, $\zeta$ are the translation and rotation friction coefficients, $\sigma$ the colloidal diameter, and $\mu_t$ and $\mu_r$ the surface mobilities determining the translational and rotational responses of the colloids to the $C$-gradient. %
The torque tends to align or anti-align the Janus particle $i$ towards particle $j$: it only has a $z$-component since the system is in 2D, and can be expressed using the relative positional angle $\chi_{ji} = \angle(\hat{\mathbf{e}}_x,\mathbf{r}_j - \mathbf{r}_i)$ as $T_{ji}(\mathbf{r}_i,\mathbf{r}_j, \theta_i)=T_r(|\mathbf{r}_j-\mathbf{r}_i|) \sin(\theta_i - \chi_{ji})$. Since only the effective shape of the cross-phoretic interactions will affect the dynamics of the particles, their origin is not relevant. The total force $\mathbf{F}_{ij}(r_{ij})$ and torque $\mathbf{T}_{ij}(r_{ij},\theta_j)$ can therefore be of different origin or include several contributions, like additional body forces.

The overdamped Langevin equations for the positions and orientations then read
\begin{align}
    \dot{\mathbf{r}}_i &= \sqrt{2D_t} \pmb{\xi}_{r,i} - \frac{k}{\gamma} (\mathbf{r}_i-\mathbf{r}_{t,i}) + v_0 \hat{\mathbf{n}}_i -\frac{F(|\mathbf{r}_j-\mathbf{r}_i|)}{\gamma} \frac{\mathbf{r}_j-\mathbf{r}_i}{|\mathbf{r}_j-\mathbf{r}_i|} \label{eq:r_sde}\\
    \dot{\theta}_i &= \sqrt{2D_r} \xi_{\theta,i} + \frac{T_r(|\mathbf{r}_j-\mathbf{r}_i|)}{\zeta} \sin(\theta_i - \chi_{ji}) \label{eq:theta_sde}
\end{align}
where $\mathbf{r}_i$ is the position of particle $i$, $D_t$ and $D_r$ the translational and rotational diffusion coefficients, $\xi$ zero-mean unit variance Gaussian white noises, and $\mathbf{r}_{t,i}$ the position of the harmonic trap for particle $i$. 

Time can be rescaled using the persistence time $\tau=1/D_r$, while distances can be rescaled by the diameter of the particles $\sigma$. Activity can be described by the Péclet number $\text{Pe}=v_0\tau/\sigma$, the stiffness of the trap by $k'=k\tau/\gamma$, and the translational diffusion coefficient by its adimensional counterpart $D_t'=D_t \tau/\sigma^2$ (arbitrarily kept as $1/3$ in all the following results). %
Brownian dynamics simulation results were obtained by applying the Euler-Maruyama method for the stochastic differential equations~\eqref{eq:r_sde} and~\eqref{eq:theta_sde}.

\section{Results} \label{sec:results}

\subsection{Single ABP in harmonic trap}
We first derive the probability distribution for a single non-interacting ABP in the harmonic trap, \emph{i.e.}, Eqs.~\eqref{eq:r_sde} and \eqref{eq:theta_sde} with $F=0$ and $T_r=0$. %
Different results have been demonstrated in similar setups~\cite{ten_hagen_brownian_2011,buttinoni_active_2022,pototsky_active_2012,malakar_steady_2020,chaudhuri_active_2021,caraglio_analytic_2022}, but they rely on the radial symmetry of the problem or use the Fokker-Planck equation. Both approaches break down when considering non-linear torques. We introduce another way to describe the steady-state probability distribution of an ABP in a stiff harmonic trap that can easily be extended to include additional interactions. This result is exact for a passive particle and asymptotically valid for an active particle with $\tau \gg \gamma/k$.

The dynamics of the orientation alone are known exactly, since it is only diffusing. We especially know that the steady-state probability distribution is uniform and given by
\begin{align}
    P(\theta) &= 1/2\pi. \label{eq:ptheta_abp}
\end{align}
When the persistence time of the ABP is significantly greater than the characteristic time of the trap ($k\tau/\gamma \gg 1$), the dynamics of the ABP can be simplified. The ABP will keep its orientation and climb up the trap potential before it reorients significantly. This displacement from the trap centre, defined by balancing the active force with the harmonic force, is given by $\gamma v_0/k$. The ABP will continuously self-propel against the restoring force of the trap, and when its orientation diffuses over long times, will move along a circular path.

\begin{figure}[t]
    \centering
    \includegraphics[width=\linewidth]{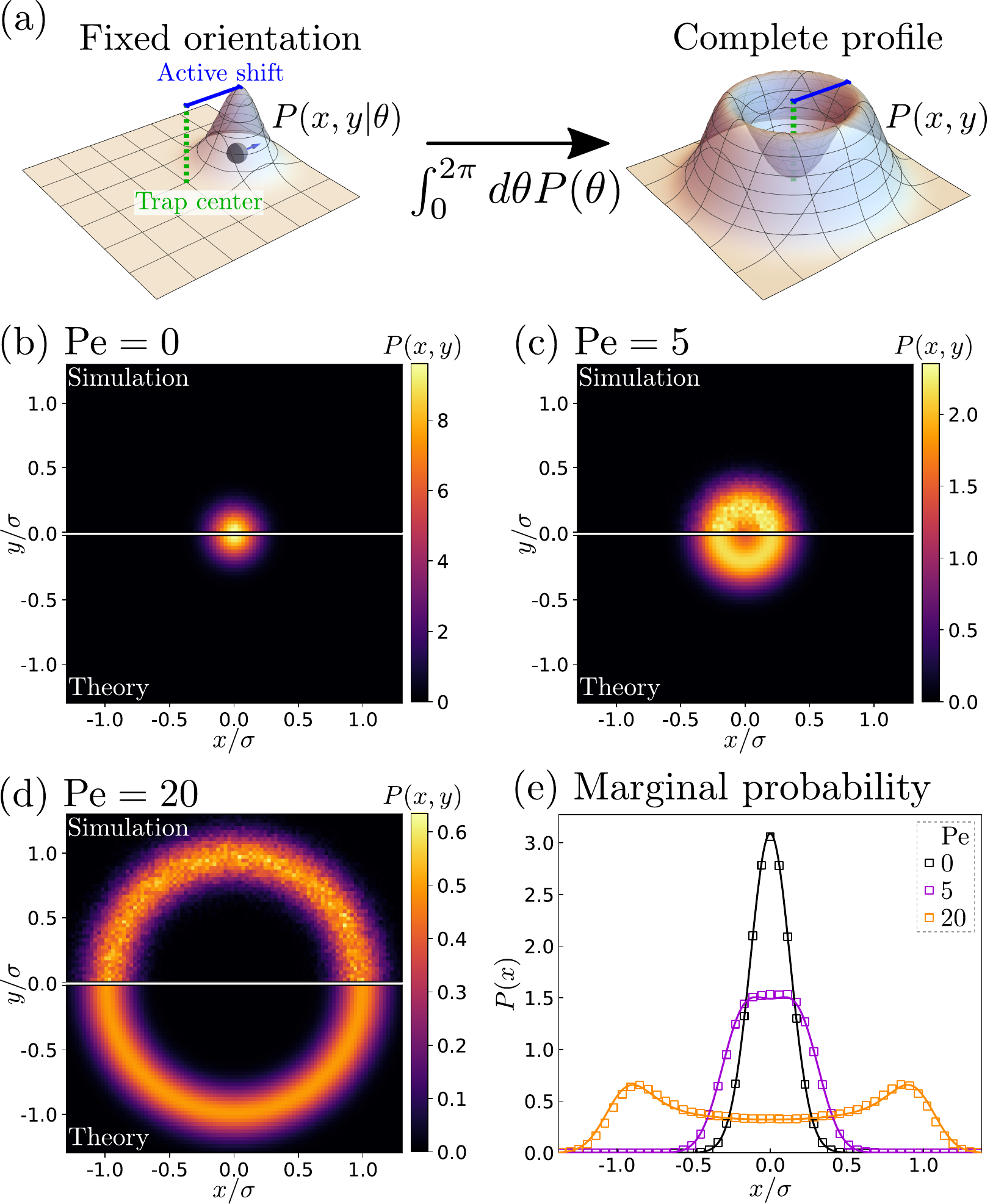}
    \caption{(a) Visualisation of the approximation required to obtain eq.~\eqref{eq:pxy_abp}. If the orientation of the ABP is locked, its probability distribution will be of a passive particle shifted from the center due to the activity. By integrating over all orientations, we recover the 2D probability distribution of an ABP in a harmonic trap. (b-d) Simulation 2D probability distributions of an ABP in a harmonic trap with $k\tau/\gamma = 20$ and varying activity $\text{Pe}$, compared with theoretical results in eq.~\eqref{eq:pxy_abp}. (e) Marginal probability distribution from the same simulations, with solids lines obtained from eq.~\eqref{eq:px_abp}.}
    \label{fig:3}
\end{figure}

We can describe the ABP by considering that for times $t\ll\tau$, it can be considered as a passive particle subject to a constant force $\gamma v_0 \hat{\mathbf{n}}_i$. The steady-state probability distribution of such a system is given by
\begin{align}
    P(x, y|\theta) &\approx \frac{k}{2\pi D_t \gamma} e^{-\frac{k}{2 D_t \gamma} \left( \left(x-\frac{\gamma v_0 \cos\theta}{k}\right)^2 + \left(y-\frac{\gamma v_0 \sin\theta}{k}\right)^2 \right)}. \label{eq:pxytheta_abp}
\end{align}
Since the ABP rotates slowly compared to its translational motion, the total probability for the orientation and the position can be approximated as $P(x,y,\theta) \approx P(x,y|\theta) P(\theta)$.
The steady-state probability distribution is then obtained by integrating over all possible orientations of the active force given in Eq.~\eqref{eq:ptheta_abp}, as illustrated in Fig.~\ref{fig:3}a, and is expressed by
\begin{align}
    P(x,y) &\approx \int_{-\pi}^\pi d\theta P(x,y| \theta) P(\theta). \label{eq:pxy}
\end{align}
Eq.~\eqref{eq:pxy} can be analytically determined using Eqs.~\eqref{eq:ptheta_abp} and~\eqref{eq:pxytheta_abp} by using the Jacobi-Anger expansion. We finally obtain
\begin{align}
    P(x,y) &\approx \frac{k}{2\pi D_t \gamma} e^{-\frac{k}{2 D_t \gamma} \left(x^2 + y^2 + (\frac{\gamma v_0}{k})^2\right)} B(x,y) \label{eq:pxy_abp} \\
    B(x,y) &= I_0\left(\frac{v_0 x}{D_t}\right) I_0\left(\frac{v_0 y}{D_t}\right) \nonumber\\
    &\quad+ 2 \sum_{i=1}^\infty (-1)^i I_{2i}\left(\frac{v_0 x}{D_t}\right) I_{2i}\left(\frac{v_0 y}{D_t}\right)
\end{align}
with $I_n$ the modified Bessel function of the first kind of order $n$.

Similar calculations can be done to obtain the marginal probability distribution only in one dimension $x$ by using the 1D probability distribution $P(x|\theta)=\int dy P(x,y|\theta)$ instead of $P(x,y|\theta)$, and we obtain
\begin{align}
    P(x) &\approx \sqrt{\frac{k}{2\pi D_t \gamma}} e^{-\frac{k}{2 D_t \gamma} \left(x^2 + \frac{1}{2}(\frac{\gamma v_0}{k})^2\right)} b(x) \label{eq:px_abp} \\
    b(x) &= I_0\left(\frac{v_0 x}{D_t}\right) I_0\left(\frac{-\gamma v_0^2}{4 D_t k}\right) + 2 \sum_{i=1}^\infty I_{2i}\left(\frac{v_0 x}{D_t}\right) I_i\left(\frac{-\gamma v_0^2}{4 D_t k}\right).
\end{align}

Examples of the results of Eqs.~\eqref{eq:pxy_abp} and~\eqref{eq:px_abp} compared to simulations are shown in Fig.~\ref{fig:3}. The analytical 2D and marginal probability distributions correctly predict the simulation results, with the Gaussian profile of the passive particle widening and becoming a ring in 2D or bimodal for the marginal one. We can also notice that we recover the exact solution of a passive particle if $v_0=0$ in Eq.~\eqref{eq:pxy_abp} and~\eqref{eq:px_abp}.

\subsection{Phoretically interacting trapped active particles}

Once we consider interacting particles, analytical results are harder to obtain. %
A single ABP in a harmonic trap moves with a characteristic distance of $\delta=2\gamma v_0 / k + \sqrt{D_t \gamma/k}$ due to activity and translational diffusion. This distance $\delta$ is small compared to the distance between the traps $D=|\mathbf{r}_{t,i} - \mathbf{r}_{t,j}|$, since $k$ is large to satisfy the persistence condition for the ABP and $D$ is large to satisfy the far-field monopole approximation. Both conditions can be easily satisfied experimentally. %
Thus, $\delta \ll D$, such that small changes in the particle position barely change the perceived pair-interactions. %
We thus approximate the interaction force and torque with the average value of the distance between the particles, $F(|\mathbf{r}_j-\mathbf{r}_i|) \approx F(d)$, $T_r(|\mathbf{r}_j-\mathbf{r}_i|) \approx T_r(d)$. We then regard the force as a constant vector directed along the direction connecting the two trap centres along the $x$ axis, while the relative positional angle $\chi_{ij}$ by symmetry becomes simply $0$ or $\pi$. %
Small variations in the particle position $x$ and $y$, impacting the varying distance between the particles $r=\sqrt{(d+\delta_x)^2+\delta_y^2}$, approximately behave as $r \approx d + \delta_x$ to first order, such that $d\gg\delta$ ensures the constants $F(d)$ and $T_r(d)$ to yield asymptotically exact interactions. %
We can thus rewrite the equations as
\begin{align}
    \dot{x}_i &= \sqrt{2D_t} \xi_{x,i} - \frac{k}{\gamma} (x_i \pm D/2) + v_0 \cos\theta_i \mp \frac{F(d)}{\gamma} \label{eq:xdot_phor}\\
    \dot{y}_i &= \sqrt{2D_t} \xi_{y,i} - \frac{k}{\gamma} y_i + v_0 \sin\theta_i\\
    \dot{\theta}_i &= \sqrt{2D_r} \xi_{\theta,i} \pm \frac{T_r(d)}{\zeta} \sin(\theta_i) \label{eq:thetadot_phor}
\end{align}
The total interaction depends now on two constants $F(d)$ and $T_r(d)$ for the force and torque respectively.

Since the equations are symmetric for the two particles, we focus on $i=2$ and suppress the index. While the average position of the particle alongside the $y$ axis is 0 by symmetry, the average position on the $x$ axis is affected by the position of the trap $D/2$, the activity $v_0$, and the interaction force $F(d)$. The particles will directly shift their position through $F(d)$, and indirectly through the torque since it breaks the radial symmetry of the activity and thus leads to asymmetric contributions through the activity $v_0$. We obtain the average position using Eq.~\eqref{eq:xdot_phor}
\begin{align}
    \langle x \rangle &= D/2 + \frac{F(d)}{k} + \frac{\gamma v_0}{k} \langle\cos\theta\rangle. \label{eq:avx_cos_phor}
\end{align}

Compared to a single ABP in a harmonic trap, the orientation is not uniform anymore because of the torque and $\langle \cos\theta\rangle \neq 0$. Since the evolution of the orientation in Eq.~\eqref{eq:thetadot_phor} is subject to thermal noise and a constant potential term, we can express the orientational probability distribution, obtain the average value of $\cos\theta$ and the average position depending on $F(d)$ and $T_r(d)$
\begin{align}
    P(\theta) &= \frac{\exp\left(\frac{T_r(d)}{\zeta D_r} \cos\theta \right)}{2\pi I_0(T_r(d)/\zeta D_r)} \label{eq:ptheta_phor}\\
    \langle\cos\theta\rangle &= \int_{-\pi}^\pi d\theta \cos(\theta) P(\theta) = \frac{I_1(T_r(d)/\zeta D_r)}{I_0(T_r(d)/\zeta D_r)} \label{eq:avcos_phor}\\
    \langle x \rangle &= D/2 + \frac{F(d)}{k} + \frac{\gamma v_0}{k} \frac{I_1(T_r(d)/\zeta D_r)}{I_0(T_r(d)/\zeta D_r)}. \label{eq:avx_phor}
\end{align}

We can define the distance $d$ between the particles from their average positions, in general defined as $d=|\langle x_2 \rangle - \langle x_1 \rangle|$. Since we are studying completely identical particles in identical traps, this can be simplified as $d = 2|\langle x \rangle|$. %
If the functional dependence of $F(r)$ and $T_r(r)$ are given, Eq.~\eqref{eq:avx_phor} can be employed to predict the average spacing between two phoretically interacting Janus particles. %

\subsection{Phoretic interactions skew the density profiles}

\begin{figure*}[t]
    \includegraphics[width=\linewidth]{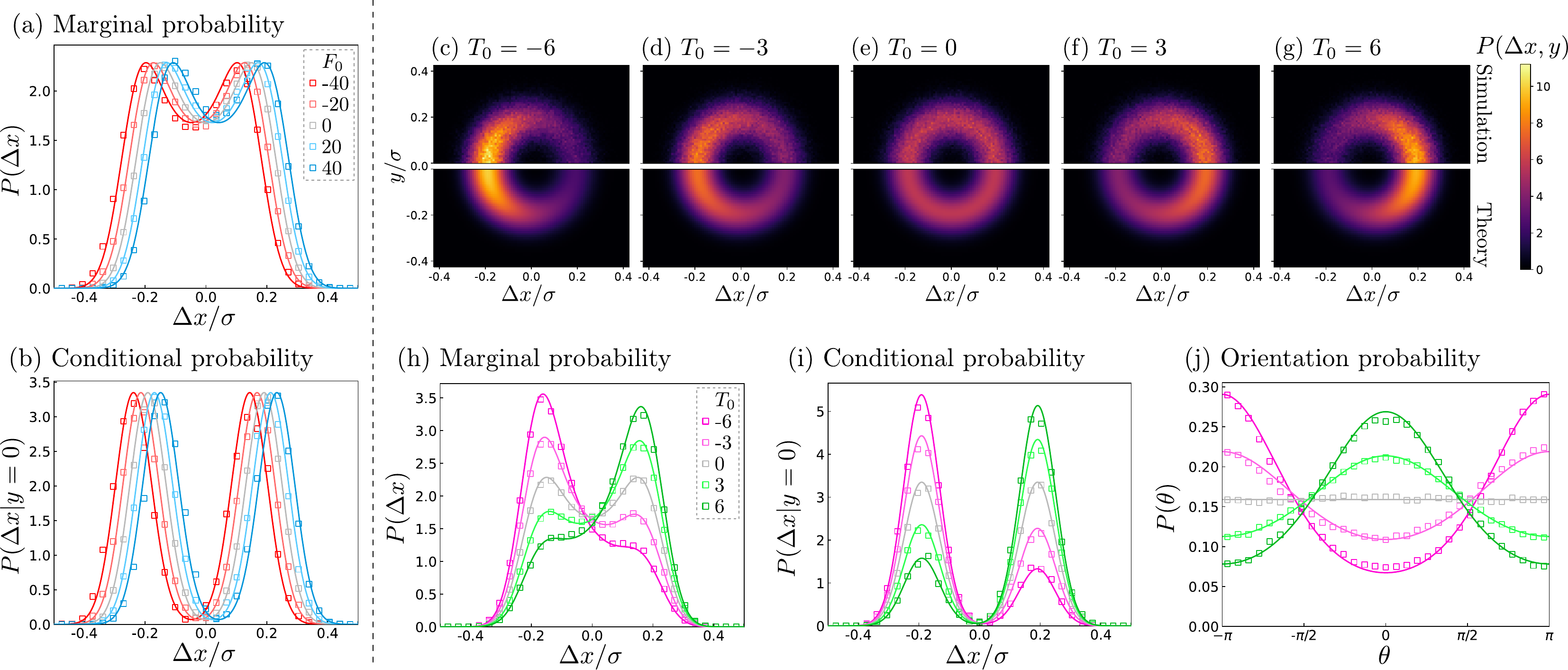}
    \caption{Effects of interactions in a system with fixed $\text{Pe}=20$, $k\tau/\gamma=100$ and $D=3$, using Brownian dynamics simulations of eqs.~\eqref{eq:r_sde} and~\eqref{eq:theta_sde}. $\Delta x = x-D/2$ is the position relative to the center of the trap. (a,b) Effects on the ABP in the right trap of the radial force between the particles, with $F(r)\frac{\tau}{\gamma \sigma} = F_0 (r/\sigma)^{-2}$ and a fixed distance between traps $D/\sigma=3$. (a) Marginal probability distribution and (b) conditional probability distribution if $y=0$ for different values of attractive and repulsive forces. (c-j) Effects on the ABP in the right trap of the torque between the particles, with the radial part $T_r(r)\frac{\tau}{\zeta} = T_0 (r/\sigma)^{-2}$ and a fixed distance between traps $D/\sigma=3$. (c-g) 2D probability distributions for 5 different prefactors describing the radial part of the torque, with the theoretical profile obtained from eq.~\eqref{eq:pxy_phor}. (h) Marginal probability distribution, (i) conditional probability distribution if $y=0$ and (j) orientation probability distributions for different values of effectively attractive and repulsive torques. Squares represent simulation results while solid lines represent theoretical results from eqs.~\eqref{eq:ptheta_phor},~\eqref{eq:px_phor} and~\eqref{eq:pxy0_phor}.}
    \label{fig:4}
\end{figure*}

The radial force between the particles shifts $P(x,y|\theta)$ in Eq.~\eqref{eq:pxytheta_abp} alongside $x$ by $F(d)/k$, whereas the torque will break the uniformity of $P(\theta)$ as expressed in Eq.~\eqref{eq:ptheta_phor}. We then obtain both the 2D and marginal probability distributions in a similar way as for a single ABP using Eq.~\eqref{eq:pxy}:
\begin{align}
    P(x,y) &\approx \frac{k e^{-\frac{k}{2 D_t \gamma} \left((x-D/2-\frac{F(d)}{k})^2 + y^2 + (\frac{\gamma v_0}{k})^2\right)}}{2\pi D_t \gamma I_0(T_r(d)/\zeta D_r)}B(x,y) \label{eq:pxy_phor}\\
    \begin{split}
        B(x,y) &= I_0\left(\frac{T_r(d)}{\zeta D_r} + \frac{v_0 (x-D/2+\frac{F(d)}{k})}{D_t}\right) I_0\left(\frac{v_0 y}{D_t}\right)\\
        +& 2 \sum_{i=1}^\infty (-1)^i I_{2i}\left(\frac{T_r(d)}{\zeta D_r} + \frac{v_0 (x-D/2-\frac{F(d)}{k)}}{D_t}\right) I_{2i}\left(\frac{v_0 y}{D_t}\right) \label{eq:bxy_phor}
    \end{split}
\end{align}
\begin{align}
    P(x) &\approx \sqrt{\frac{k}{2\pi D_t \zeta}} \frac{e^{-\frac{k}{2 D_t \zeta} \left((x-D/2-\frac{F(d)}{k})^2 + \frac{1}{2}(\frac{\gamma v_0}{k})^2\right)}}{2\pi I_0(T_r(d)/ \zeta D_r)} b(x) \label{eq:px_phor}\\ 
    \begin{split}
        b(x) &= I_0\left(\frac{T_r(d)}{\zeta D_r} + \frac{v_0 (x-D/2-\frac{F(d)}{k})}{D_t}\right) I_0\left(\frac{-\gamma v_0^2}{4 D_t k} \right)
        \\ +& 2 \sum_{i=1}^\infty I_{2i}\left(\frac{T_r(d)}{\zeta D_r} + \frac{v_0 (x-D/2-\frac{F(d)}{k})}{D_t}\right) I_i\left(\frac{-\gamma v_0^2}{4 D_t k} \right). \label{eq:bx_phor}
    \end{split}
\end{align}
Despite the fact that these expression give accurate results, they are increasingly harder to compute when increasing $v_0/k$. This is due to the fact that more terms of the sum need to be taken into account before it converges. %
Since $I_{2n}(0)=0$ for $n=1,2,3,\dots$, we note that the expression of $B(x,y)$ in Eq.~\eqref{eq:bxy_phor} greatly simplifies if $y=0$, and we obtain the simpler expression
\begin{align}
    \begin{split}
        P(x|y=0) &\approx c e^{-\frac{k}{2 D_t \gamma} \left((x-D/2-\frac{F(d)}{k})^2 + (\frac{\gamma v_0}{k})^2\right)}\\
        \times& \ I_0\left(\frac{T_r(d)}{\zeta D_r} + \frac{v_0 (x-D/2+\frac{F(d)}{k})}{D_t}\right)
    \end{split}
    \label{eq:pxy0_phor}
\end{align}
with the constant $c$ defined through the normalisation $\int dx P(x|y=0)=1$. This conditional probability distribution immediately converges while encapsulating all the important effects of the interactions on a particle, making it a more reliable quantity to study the system systematically.

Using these expressions, we show their validity compared to simulations of Eqs.~\eqref{eq:r_sde} and \eqref{eq:theta_sde} and understand the effects of the force and torque separately in Fig~\ref{fig:4}. If the particles are interacting through a radial force $F(r)\frac{\tau}{\gamma \sigma} = F_0 (r/\sigma)^{-2}$, with $F_0$ the adimensional strength of the force, we observe that the density profile shape does not change, and is only shifted by $F(d)/k$, as observed for passive particle~\cite{buttinoni_mechanical_2021}. We also see that the main difference between the marginal (Fig.~\ref{fig:4}a) and conditional (Fig.~\ref{fig:4}b) probability distributions is that the two peaks are clearly separated in the second case. This is due to the fact that the main contribution to the conditional probability distribution is when the orientation is almost aligned with the $x$ axis, whereas the $x$ position of the particle contributes to the marginal one regardless of the orientation.

However, when the interactions between the particles induce an effective torque with a radial component $T_r(r)\frac{\tau}{\zeta}=T_0 (r/\sigma)^{-2}$, with $T_0$ the adimensional strength of the torque, we clearly see that this additional contribution breaks the circular symmetry observed in a single ABP in a harmonic trap. The particle has a tendency to face towards (opposite of) the other trap if $T_r(d)<0$ ($T_r(d)>0$), leading to an effective attraction (repulsion). The direct effect is to break the uniformity of the density profile of the orientation (Fig.~\ref{fig:4}j), inducing an indirect change of relative scale of the peaks of the bimodal profiles (Fig.~\ref{fig:4}h,i) through the activity. This can be seen in the 2D density profiles (Fig.~\ref{fig:4}c-g), where the x symmetry is broken while the y symmetry is intact. One can also notice that the effect of the torque is asymmetric for opposite values of $F_0$. This occurs because the average value $T_r(d)$ determines the effect on the probability distribution, where $d$ is also affected by the torque. An effectively attractive torque will reduce the distance $d$, increasing $T_r(d)$ compared to an effectively repulsive torque that would increase $d$. This also happens for the force in Fig.~\ref{fig:4}a,b, albeit less noticeable visually.

\subsection{Unique phoretic signatures of positional moments}

Having described the steady-state properties of these interacting active particles, we now reverse the reasoning, and infer the pair-interactions between them only from the steady-state information that we can observe experimentally. One way of doing so is to study how the moments of the steady-state probability distributions depend on the torque and force, as presented in this section. From the interactions, it is then possible to understand which physical phenomena dominate.

\begin{figure}[t]
    \includegraphics[width=\linewidth]{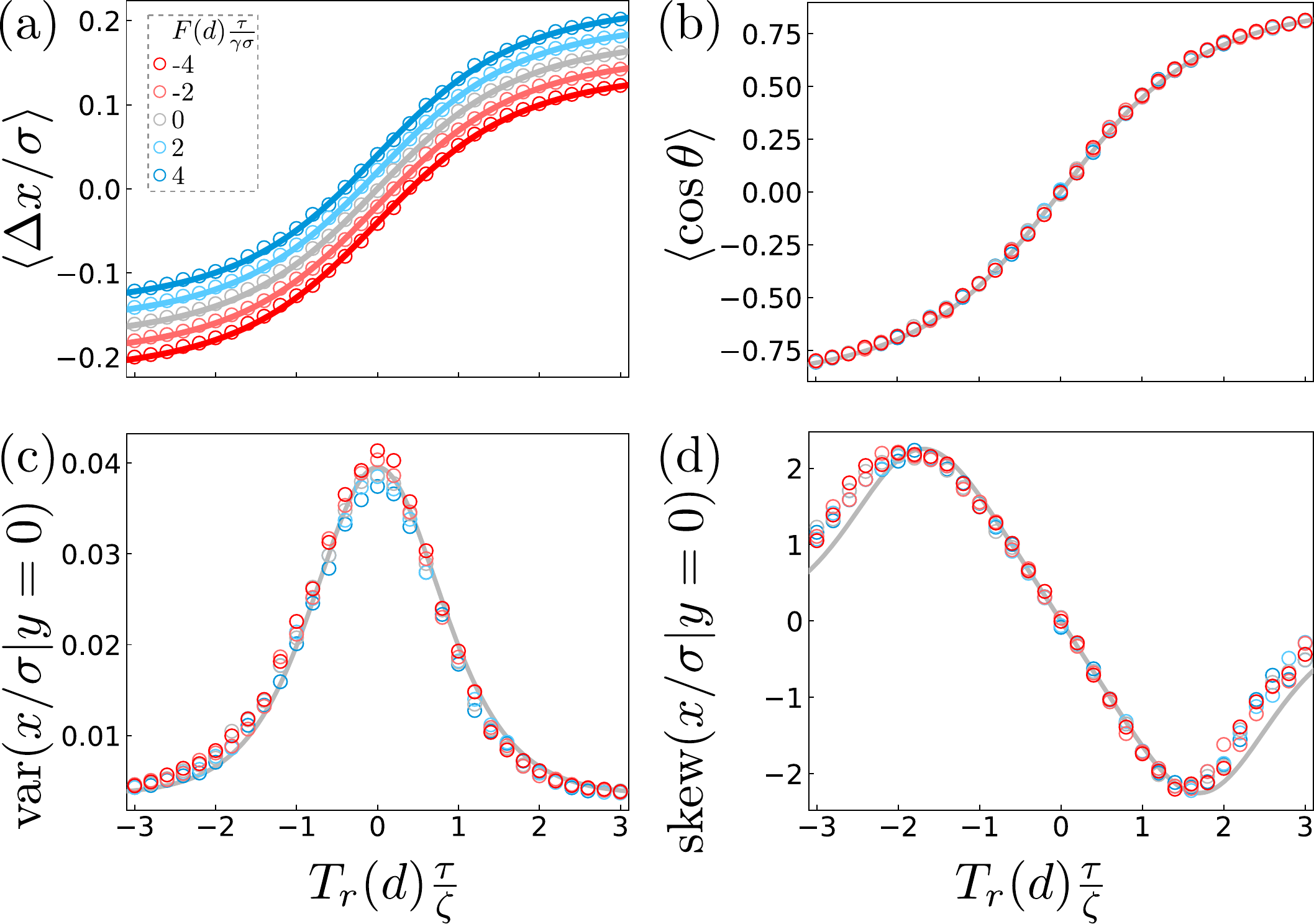}
    \caption{Various moments of the probability distributions of the ABP phoretically interacting with another ABP on its left depending on the average torque $T_r(d)$, shown for different values of the average force $F(d)$, with a fixed distance between traps. ($D/\sigma=3$, $k\tau/\gamma=100$, $\text{Pe}=20$) (a) Average displacement inside the trap. (b) Average of $\cos\theta$. (c) Variance and (d) skewness of the probability distribution under the condition that $y=0$. (a-d) Circles represent simulation results while solid lines represent theoretical result from eqs.~\eqref{eq:avcos_phor},~\eqref{eq:avx_phor},~\eqref{eq:vary0}, and~\eqref{eq:skewy0}, with a single line for (b-d) since these moments do not depend on the force.}
    \label{fig:5}
\end{figure}

The average $x$-position and the average cosine of the orientation are given in Eqs.~\eqref{eq:avx_phor} and~\eqref{eq:avcos_phor}, and are plotted in Fig.~\ref{fig:5}a,b depending on the torque for different values of the force. While the force only affects the first moment of the position linearly, the average orientation is independent of it. The torque affects moments of both the position and the orientation non-linearly. Since $\langle x \rangle$ depends linearly on $\langle\cos\theta\rangle$, they are both affected the same way by the torque. %
Around $T_r=0$, we can use the series definition of the modified Bessel function of the first kind to show that $\langle\cos\theta\rangle \approx T_r/2$. When increasing the magnitude of the torque, $\langle\cos\theta\rangle$ increases and saturates to $T_r/|T_r|$ at infinity. We see that we continuously go from the linear response, if the torque is considered as a small perturbation, to the strong torque case where the orientation is locked into one of the emergent wells of the periodic potential in eq.~\eqref{eq:ptheta_phor} and Fig.~\ref{fig:4}.

We obtain the variance and the skewness of the profile $P(x|y=0)$ by successively and numerically integrating Eq.~\eqref{eq:pxy0_phor} to obtain the mean $m=\int_\mathbb{R} dx \, x P(x|y=0)$, then the variance and the skewness
\begin{align}
    \text{var}(x|y=0) &= \int_\mathbb{R} dx \, (x-m)^2 P(x|y=0) \label{eq:vary0}\\
    \text{skew}(x|y=0) &= \int_\mathbb{R} dx \left(\frac{x-m}{\sqrt{\text{var}(x|y=0)}}\right)^3 P(x|y=0). \label{eq:skewy0}
\end{align}
This is plotted by a single solid line in Fig.~\ref{fig:5}c,d since it does not depend on $F$, and compared to values directly computed from simulations. We see that the torque decreases the variance in Fig.~\ref{fig:5}c, because one of the peaks grows while the other shrinks, which reduces the important contributions to the variance away from the average. 
When the torque increases, the orientation becomes more and more locked in one specific direction, eventually becoming almost constant. In this case, the active particle becomes mathematically equivalent to a passive particle under a constant force, for which the variance is $\gamma D_t/k$. %
We can also have an upper boundary for the variance by approximating $P(x|y=0)$ to two Gaussian peaks of 1D passive particles for $T_r=0$, which is satisfied if $\gamma v_0 /\sigma k \gg 1$ and $k\tau/\gamma\gg1$. In this case, we can compute the variance to be $\gamma D_t/k + (\gamma v_0 / k)^2$. This approximation and upper limit become exact for large values of $v_0/k$.

The torque has a different impact on the skewness of $P(x|y=0)$, as shown in Fig.~\ref{fig:5}d. When the torque is zero, the skewness is also zero since the density profile is symmetric. An effectively attractive torque will increase the probability to be on the left side of the trap center, inducing a positive skewness, and \emph{vice versa} for a repulsive torque. However, by increasing the magnitude of the torque, we reach a point where the density profile becomes closer to a single Gaussian peak. This implies that the mean gets closer to the centre of the main Gaussian peak, and the contribution of the secondary one to the skewness gets smaller. Overall, the magnitude of the skewness now decreases, and approaches $0$ for an increasingly large magnitude of the torque.

\subsection{Reverse engineering interaction landscapes}

\begin{figure}[t]
    \includegraphics[width=\linewidth]{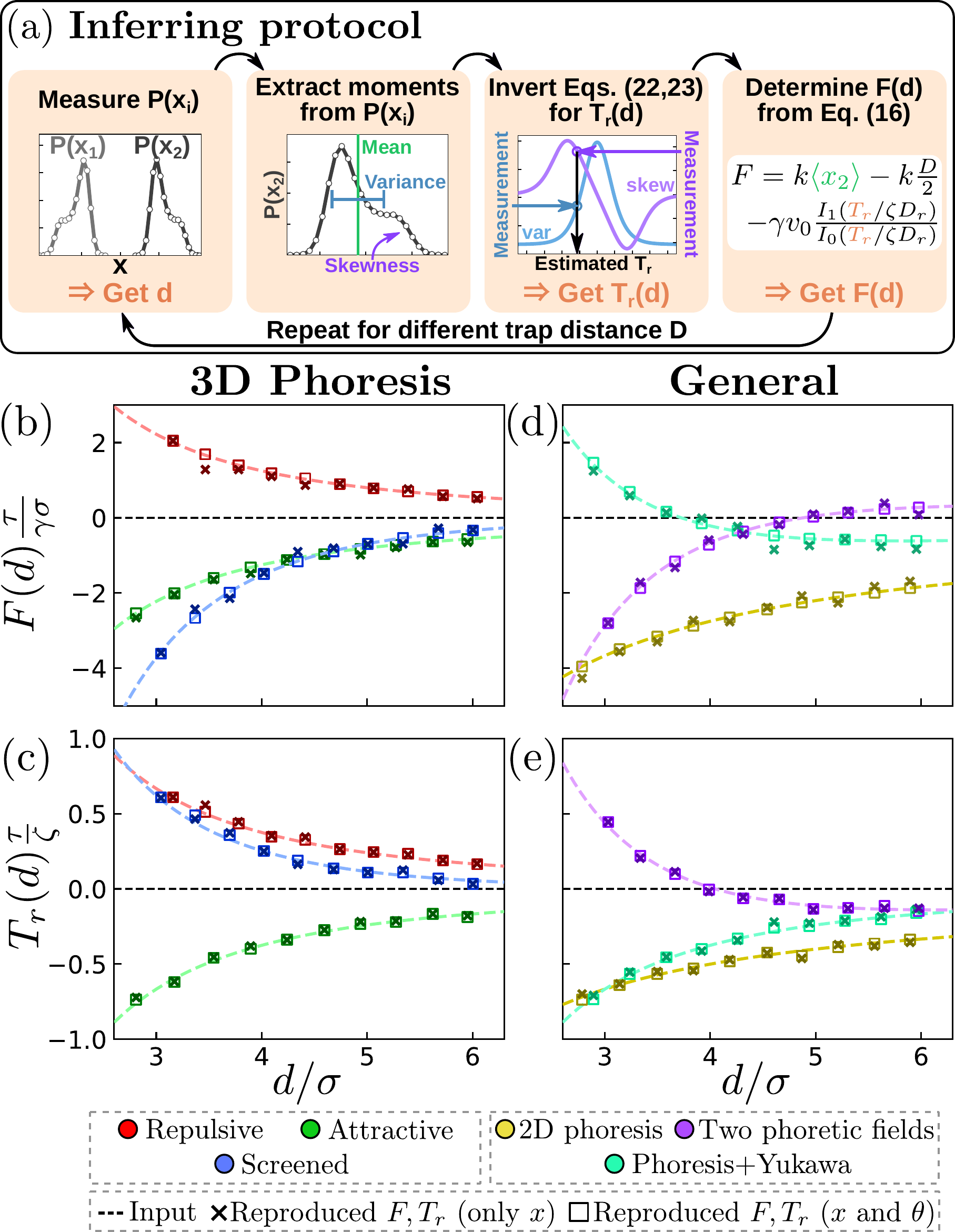}
    \caption{Inferring pair interactions. (a) Description of the protocol used to infer the force $F$ and torque $T_r$ from the measured probability distribution $P(x_i)$. Reproduced force (b,d) and torque (c,e) between the ABPs, using solely positions (crosses) or both position and orientations (squares). %
    The force and torque are $F(d)\frac{\tau}{\gamma\sigma}= F_U \frac{\sigma^2}{r^2} + F_S (\frac{\sigma^2}{r^2} + \frac{\sigma^2\kappa}{r}) e^{-\kappa r}$ and $T_r(d)\frac{\tau}{\zeta}= T_U \frac{\sigma^2}{r^2} + T_S (\frac{\sigma^2}{r^2} + \frac{\sigma^2\kappa}{r}) e^{-\kappa r}$ with $F_U, F_S, T_U, T_S$ defining the interaction strength and $\kappa^{-1}$ the screening length. %
    (b,c) Standard 3D phoresis implying repulsive (red, $F_U=20, F_S=0, T_U=6, T_S=0$) or attractive (green, $F_U=-20, F_S=0, T_U=-6, T_S=0$) interactions, and a screened phoretic field (blue, $F_U=0, F_S=-60, T_U=0, T_S=10, \kappa=\sigma^{-1}/2$). %
    (d,e) Three general examples: 2D phoretic field (yellow, $F(r)\frac{\tau}{\gamma\sigma}=-11\sigma/r$ and $T_r(r)\frac{\tau}{\zeta}=-2\sigma/r$), combination of two independent phoretic fields (unscreened and screened) (purple, $F_U=30, F_S=-100, T_U=-10, T_S=25, \kappa=\sigma^{-1}/2$), combination of phoretically interacting colloids with an additional electrostatic Yukawa repulsion (cyan, $F_U=-40, F_S=90, T_U=-6, T_S=0, \kappa=\sigma^{-1}/2$).}
    \label{fig:6}
\end{figure}

With the results we have now derived, we are able to infer both the force and the torque separately. The general method to do so is described in Fig.~\ref{fig:6}a. By tracking the position of the colloids, each of their probability distributions $P(x_i)$ can be obtained. The distance $d$ between them can be directly measured, as well as the mean position $\langle x \rangle$, the variance $\text{var}(x|y=0)$ and the skewness $\text{skew}(x|y=0)$. The moments can then be used to infer the force $F(d)$ and torque $T_r(d)$ for the distance $d$. Since moments higher than the mean are solely influenced by the torque, the variance and skewness from Eqs.~\eqref{eq:vary0} and~\eqref{eq:skewy0} can be numerically inverted for $T_r(d)$ to infer its value. %
Using the measured mean position $\langle x \rangle$ and $T_r(d)$, the force $F(d)$ can now be directly determined from Eq.~\eqref{eq:avx_phor}. By repeating this procedure for different distances $D$ between the traps, the force $F$ and torque $T_r$ are estimated for different distances $d$, giving information about their intensity and spatial dependency.

We illustrate this method by performing simulations with different input for $F(r)$ and $T_r(r)$ (lines in Fig.~\ref{fig:6}b-e) which we reproduce from the trajectories (symbols in Fig.~\ref{fig:6}b-e). %
We first consider particles only interacting through bulk 3D cross-phoresis (Fig.~\ref{fig:6}b,c). The resulting force and torque can separately be attractive or repulsive depending on the physical properties of the Janus particles, and scale with $|\nabla C(\mathbf{r})| \propto r^{-2}$ or $|\nabla C(\mathbf{r})| \propto (\frac{1}{r^2} + \frac{\kappa}{r}) e^{-\kappa r}$ if the phoretic field is subject to screening. Following the proposed protocol, we can accurately reproduce the simulation input, proving its applicability. %
Note, that if the particle orientation is experimentally accessible, the expression of $\langle \cos\theta\rangle$ from Eq.~\eqref{eq:avcos_phor} can also be numerically inverted to infer the torque. In this way, the reproduced interactions do not rely on the asymptotically exact approximations underlying Eq.~\eqref{eq:pxy}, granting slightly higher accuracy.

In order to demonstrate that the proposed protocol is readily generalisable, we apply it to more general scenarios in Fig.~\ref{fig:6}d,e. %
If the colloids were held between two planar boundaries, the diffusion governing the phoretic field might effectively become 2D, resulting in a different scaling for the force and torque $|\nabla C(\mathbf{r})| \propto r^{-1}$. However, this does not affect the applicability of the method as shown by the yellow symbols in Fig.~\ref{fig:6}d,e. %
Similarly, in the case of overlapping interactions, such as two phoretic fields, thermophoresis and screened diffusiophoresis, where each independently affects the slip velocities around the colloids, or a combination of phoretic and electrostatic (Yukawa) interactions, we accurately reproduce the input, represented by the purple and cyan symbols in Fig.~\ref{fig:6}d,e, respectively. %
Our results can thus be employed to characterise diverse interaction mechanisms and subsequently provide insights into the processes underlying the interplay between active particles.

\section{Conclusion} \label{sec:conclusion}
Due to the complexity of the interactions between active Janus particles, there is a need to reliably understand how they interact to predict their behaviour when functionalised. Optical tweezers pose a reliable instrument to measure pair-interactions, as already done with passive particles~\cite{meiners_direct_1999,buttinoni_mechanical_2021}. %
In this work, we described how phoretic interactions between two active Brownian Janus particles in two separate harmonic traps affect their steady-state probability distributions and moments, and used it to infer the space-dependent pair interactions by varying the distance between the traps. Describing an ABP in a harmonic trap in the long persistence regime interacting with another colloid, we show that a phoretic force simply displaces the average position, while phoretic torques break the orientational isotropy leading to an orientational bias and consequently to a skewed positional probability distribution. We show furthermore that moments of the probability distributions can be used to infer the total force and torque induced by each particle on the other as functions of the distance $d$ between their average positions. Particularly, both torque and force can be measured from the experimentally tracked colloidal positions, required for index-matched Janus particles~\cite{buttinoni_active_2022} for which the orientation cannot be easily tracked.

Our results and methods constitute a conceptual advance for measuring pair interactions in active systems. This can be extended to two different Janus particles that would interact non-reciprocally, as can be observed experimentally~\cite{schmidt_light-controlled_2019}, and serves as a case study for any type of pair interaction. %
Once the spatial dependency of the interactions has been measured, underlying properties of the interactions can also be studied. In the case of phoresis, surface properties of the colloids or screening from the environment can be estimated from the interactions, providing valuable insights into the microscopic details of nonequilibrium colloidal physics. %
A natural extension to this work would be to incorporate hydrodynamic interactions and how they would impact the system in tandem with other interactions, which would allow us to understand the physics of Janus particles for an even broader range of systems.

\vskip 0.2 cm
\section*{Acknowledgments} 

H.L. acknowledges funding by the Deutsche Forschungsgemeinschaft (DFG) within project LO 418/29-1. I.B. acknowledges funding by the Deutsche Forschungsgemeinschaft (DFG) within project BU 4040/3-1.

\vskip 0.2 cm
\section*{Author declarations} 

Conflict of Interest: The authors have no conflicts to disclose.

\vskip 0.2 cm
\section*{Data availability} 

The data that supports the findings of this study are available from the corresponding author upon reasonable request.
 

\bibliography{InferringInteractions}

\end{document}